\documentclass[sigconf,nonacm]{acmart}

\usepackage{acmart-taps}

\usepackage{cleveref}
\usepackage{algorithm}
\usepackage{algorithmic}
\usepackage{mathtools}
\usepackage{setspace}
\usepackage{placeins}
\usepackage{listings}
\usepackage{caption}

\RequirePackage{xcolor}
\RequirePackage{tagging}

\definecolor{colsample}{gray}{0.45}
\definecolor{colexpl}{rgb}{0.4, 0.6, 0.8}

\newcounter{artifactCounter}

\usepackage{subfig}
\usepackage{amsmath,amsfonts}
\usepackage{algorithmic}
\usepackage{graphicx}
\usepackage{textcomp}
\usepackage{xcolor}
\usepackage{booktabs}
\usepackage{tagging}
\usepackage{listings}
\usepackage{xcolor}  
\usepackage{fancyhdr}
\AtBeginDocument{%
  }

\begin{document}

\title{PGT-I: Scaling Spatiotemporal GNNs with Memory-Efficient Distributed Training}

\author{Seth Ockerman}
\affiliation{
  \institution{University of Wisconsin-Madison}
  \city{Madison}
  \state{WI}
  \country{USA}
}
\additionalaffiliation{
\institution{Argonne National Laboratory}
 \city{Lemont}
 \state{IL}
 \country{USA}
}
\email{sockerman@cs.wisc.edu}

\author{Amal Gueroudji}
\affiliation{
  \institution{Argonne National Laboratory}
  \city{Lemont}
  \state{IL}
  \country{USA}
}
\email{agueroudji@anl.gov}

\author{Tanwi Mallick}
\affiliation{
  \institution{Argonne National Laboratory}
  \city{Lemont}
  \state{IL}
  \country{USA}
}
\email{tmallick@anl.gov}

\author{Yixuan He}
\affiliation{
  \institution{Arizona State University}
  \city{Phoenix}
  \state{AZ}
  \country{USA}
}
\email{Yixuan.He@asu.edu}

\author{Line Pouchard}
\affiliation{
  \institution{Sandia National Laboratories}
  \city{Albuquerque}
  \state{NM}
  \country{USA}
}
\email{lcpouch@sandia.gov}

\author{Robert Ross}
\affiliation{
  \institution{Argonne National Laboratory}
  \city{Lemont}
  \state{IL}
  \country{USA}
}
\email{rross@anl.gov}

\author{Shivaram Venkataraman}
\affiliation{
  \institution{University of Wisconsin-Madison}
  \city{Madison}
  \state{WI}
  \country{USA}
}
\email{shivaram@cs.wisc.edu}

\renewcommand{\shortauthors}{Ockerman et al.}

\begin{abstract}

Spatiotemporal graph neural networks (ST-GNNs) are powerful tools for modeling spatial and temporal data dependencies. However, their applications have  been limited primarily to small-scale datasets because of memory constraints. While distributed training offers a solution, current frameworks lack support for spatiotemporal models and overlook the properties of spatiotemporal data. Informed by a scaling study on a large-scale workload, we present PyTorch Geometric Temporal Index (PGT-I), an extension to PyTorch Geometric Temporal that integrates distributed data parallel training and two novel strategies: index-batching and distributed-index-batching. Our index techniques exploit spatiotemporal structure to construct snapshots dynamically at runtime, significantly reducing memory overhead, while distributed-index-batching extends this approach by enabling scalable processing across multiple GPUs. Our techniques enable the first-ever training of an ST-GNN on the entire PeMS dataset without graph partitioning, reducing peak memory usage by up to 89\% and achieving up to a 11.78x speedup over standard DDP with 128 GPUs. \footnote{To appear in the 2025 International Conference for High Performance Computing,
Networking, Storage, and Analysis.}\enlargethispage{12pt}
\end{abstract}


\maketitle

\section{Introduction}

Machine learning (ML) is increasingly popular because of its effectiveness, and the advent of big data has enabled applications in numerous domains such as digital agriculture~\cite {Wolfert2017Big, Meshram2021Machine}, climate modeling~\cite{Kashinath2021Physics, De2023Machine, Gorman2018Using},  and public health~\cite{Olawade2023Using, Mhasawade2021Machine}. Many of these applications rely on data that combines spatial and temporal information, necessitating models that can predict future events based on patterns that unfold across both dimensions. ST-GNNs~\cite{sahili2023spatiotemporalgraphneuralnetworks} address this challenge; these models are used for tasks such as traffic prediction~\cite{li2021dynamic,Yu2018Spatio,zuo2022graphconvolutionalnetworkstraffic}, energy modeling~\cite{Daenens2024_Spatio, Karimi2021_Spatio, Fan2024_UsingSG}, and infectious disease forecasting~\cite{Croft2023_forecasting, Fritz2022_Combining, rozemberczki2021_pytorch}. Spatiotemporal data consists of a series of graphs, where graph nodes and edges capture spatial relationships, structured as a time series that captures temporal trends. To process spatiotemporal data, ST-GNNs use a combination of recurrent~\cite{Rumelhart1987Learning, Jordan1986Serial} and convolutional~\cite{Lecun1998} layers, enabling them to model complex data dependencies and make precise predictions. \enlargethispage{12pt}

There is a growing need to support training on larger spatiotemporal datasets to increase model accuracy~\cite{Mallick2020Graph,guan2022dynagraph,Wang2024Bricks,kong2025graphsparsenetnovelmethodlarge} and enable large-scale prediction for tasks such as statewide traffic prediction and detailed epidemiological modeling. Currently, the majority of ST-GNN applications have been limited to small to moderate-sized datasets~\cite{Yu2018Spatio, Wu2019_Graph, Bai2020_Adaptive, Li2023_self}; existing ST-GNN tools~\cite{rozemberczki2021_pytorch,li2018dcrnn_traffic,li2021dynamic} support only single-GPU computation and impose substantial memory requirements, which can exceed the memory capacity of even state-of-the-art compute nodes. For example, PeMS, a modest 8~GB dataset before preprocessing, crashes because of out-of-memory (OOM) errors on a compute node with 512~GB of RAM. Furthermore, even for smaller datasets that fit within system memory, GPU acceleration is limited to individual batches, and the broader workflow remains bottlenecked by frequent CPU-to-GPU memory transfers. While distributed computation presents a potential solution, existing distributed training tools~\cite{Zhou2023_DistTGL,Swapnil2021_P3,wang2020deep,Zheng2020Dist} do not support many of the necessary operations for ST-GNN models, lack compatibility with spatiotemporal data, and limit optimization to model training. These limitations prevent ST-GNN workflows from taking full advantage of a high-performance computing (HPC) environment's distributed infrastructure and powerful GPUs.

In this work we investigate how to increase the efficiency of ST-GNN training and best utilize modern HPC machines, which include state-of-the-art GPU accelerators and distributed infrastructures. We propose a set of techniques designed to reduce the memory demands of spatiotemporal data, reduce overhead due to frequent CPU-to-GPU memory transfers, and enable distributed data parallel (DDP) training with spatiotemporal models. We begin by analyzing model training on the PeMS-All-LA dataset -- noted for its memory and runtime challenges in prior work~\cite{Mallick2020Graph} -- to identify key obstacles to scaling ST-GNNs. Our analysis uses both a baseline PyTorch implementation of the popular Diffusion Convolutional Recurrent Neural Network (DCRNN)~\cite{li2018dcrnn_traffic} and an optimized PGT~\cite{rozemberczki2021_pytorch} implementation. We identify that existing tools are unable to scale to large datasets because of significant memory overhead caused by data duplication during preprocessing and the lack of a distributed implementation.


 Based on our analysis, we propose a new index-based approach to preprocessing and training. Index-batching reorganizes spatiotemporal preprocessing to eliminate unneeded memory duplication and constructs dataset items at runtime with low overhead. We utilize index-batching's low memory footprint to enable GPU-index-batching, a technique that performs the ST-GNN workflow (i.e., preprocessing and training) entirely within GPU memory and consolidates CPU-to-GPU memory transfers to a single operation at the beginning of preprocessing. To enable full use of the distributed infrastructure of HPC environments, we combine our optimizations with DDP training as distributed-index-batching. Our techniques allow us to scale training, for the first time, to the full PeMS dataset without graph partitioning. Notably, our techniques are applicable to any model that operates on spatiotemporal data in a sequence-to-sequence format, enabling their adoption across a broad range of models (e.g., ST-LLMs \cite{liu2024_spatial}, ST-GNNs \cite{shleifer2019incrementallyimprovinggraphwavenet, li2021dynamic}, and attention-based-GNNs \cite{Guo_Lin_Feng_Song_Wan_2019,zhu2020a3tgcnattentiontemporalgraph,grigsby2023longrangetransformersdynamicspatiotemporal}). We publish our work as \texttt{PyTorch Geometric Temporal Index}, an extension to PGT that incorporates DDP training and is specifically designed for spatiotemporal training. In summary, we make the following major contributions: 

\begin{itemize}
    \item We propose and evaluate two novel techniques — index-batching and distributed-index-batching — and their respective GPU optimizations that increase memory efficiency and scalability for the spatiotemporal family of models.
    
    \item We present the first open-source DDP training framework tailored for ST-GNNs, integrating our techniques into PGT\footnote{\url{https://github.com/benedekrozemberczki/pytorch_geometric_temporal}} with minimal changes to the existing workflow.
    
    \item We validate PGT-I's efficacy through a scaling study with up to 128 GPUs on ALCF's Polaris supercomputer ~\footnote{\url{https://www.alcf.anl.gov/polaris}} using the PeMS dataset and DCRNN. Our results demonstrate that our techniques scale efficiently, achieving up to a 115.49x reduction in training time with 128 GPUs, while also outperforming standard DDP training by up to 11.78x in terms of workflow runtime and reducing memory usage by up to 89\%.
\end{itemize}

\section{Background}
This work focuses on the distributed training of ST-GNNs. Accordingly, \cref{sec:brk-data} provides an introduction to spatiotemporal data, \cref{sec:bkr-stGNN} introduces ST-GNNs and DCRNN, and \cref{sec:bkr-pre} describes ST-GNN data preprocessing. To provide clarity, we define the terminology relating to ``nodes" as follows. ``Graph node" refers to an individual entity within a network in the context of graph theory. For example, we may refer to a ``graph with 300 nodes" to define the number of nodes within its network. The other use of ``node" in this paper is ``compute node." In this work, ``compute node" refers to an individual computational unit within the context of a larger HPC environment.

\subsection{Spatiotemporal Data}
\label{sec:brk-data}
Spatiotemporal data records the evolution of features across both space and time, integrating location with temporal progression. At a high level, each spatiotemporal dataset captures a record of features over time as a series of graphs. Each time step in the dataset corresponds to a snapshot of the graph, where nodes are associated with feature vectors that change over time -- such as sensor readings, traffic speeds, or environmental metrics. The temporal dimension adds sequential context, capturing how node features evolve, while the spatial structure models interactions between different locations. Together, this structure enables models to reason about patterns that evolve along both dimensions. To encode spatial information as graph nodes, practitioners load information about node IDs and their corresponding latitudes and longitudes from a separate file as an adjacency matrix. A simple transformation can be applied to the adjacency matrix to generate a weighted matrix that encodes the strength of connections between graph nodes.

\begin{table*}[h]
\centering
\scalebox{0.90}{ 
\begin{tabular}{|c|c|c|c|c|c|c|}
\hline
\textbf{Dataset} & \textbf{Type} & \textbf{Features} &  \textbf{Nodes} & \textbf{Entries} & \textbf{Size Before Preprocessing} & \textbf{Size After Preprocessing} \\
\hline
Chickenpox-Hungary & Epidemiological   & case count & 20 & 522 & 83.36~KB & 657.92~KB   \\
\hline
Windmill-Large & Energy & hourly energy output & 319 & 17,472 & 44.59~MB & 712.80~MB   \\
\hline
METR-LA &  Traffic & speed, day of week & 207 & 34,272 & 54.39~MB & 2.54~GB   \\
\hline
PeMS-BAY &  Traffic & speed, day of week & 325 & 52,105 & 129.62~MB & 6.05~GB   \\
\hline
PeMS-All-LA & Traffic & speed, day of week & 2716 & 105,120 & 2.12~GB & 102.08~GB   \\
\hline
PeMS &  Traffic & speed, day of week & 11,160 & 105,120 &  8.71~GB & 419.46~GB   \\
\hline
\end{tabular}
}
\vspace{5pt}
\caption{Summary of the datasets used and their sizes before and after preprocessing with float64 precision. Datasets are listed in ascending order of size.}
\label{tab:bkrnd-datasets}\vspace*{-9pt}
\end{table*}

\subsection{Spatiotemporal Graph Neural Networks}
\label{sec:bkr-stGNN}
ST-GNNs are a specialized type of graph neural network (GNN) designed to model and analyze spatiotemporal data. They differ from standard GNNs and temporal graph neural networks (T-GNNs) in a few key ways. GNNs operate on graph-structured data, predicting the properties of graph edges and nodes, which are typically static. In contrast, ST-GNNs train on graph data that evolves over time. T-GNNs train on graphs with dynamically evolving topology or features, focusing on temporal dependencies; however, T-GNNs do not explicitly model spatial correlations, which are central to spatiotemporal training that jointly learns spatial and temporal dependencies over a static or evolving graph.

ST-GNNs, such as DCRNN described in~\citet{li2018dcrnn_traffic}, use a graph-based encoder-decoder architecture to capture the spatial and temporal patterns in data. In this work we focus on spatiotemporal data of three primary types: epidemiological (disease spread), energy modeling, and traffic prediction. As described in \cref{sec:brk-data}, the connection data can be modeled as a static graph \( G = (V, E, A) \), where \( V \) is a set of \( N \) graph nodes representing the locations, \( E \) is a set of directed edges representing the connections between graph nodes, and \( A \in \mathbb{R}^{N \times N} \) is the weighted adjacency matrix representing the strength of proximity between graph nodes. We adopt a static graph representation with dynamic/temporal signal as defined in ~\citet{rozemberczki2021_pytorch}. Given the historical observations/signals $X(t-T'+1), \dots, X(t)$ at each node of the graph, the goal is to learn a function \( f(\cdot) \) that takes observations for \( T' \) time steps as input to forecast the relevant metric (e.g., chickenpox cases in a city) for the next \( T \) time steps: 
\[
X(t-T'+1), \ldots, X(t); G \xrightarrow{f(\cdot)} X(t+1), \ldots, X(t+T).
\]

Using graph-structured time series data and the corresponding adjacency matrix, DCRNN performs graph convolutions within a recurrent neural network. To model spatial relationships within the graph, each layer calculates a respective graph node's value, also known as its feature, by aggregating the features of its spatial neighbors. Layers capture the temporal aspect of the data by incorporating the gated mechanisms of recurrent neural networks, which retain the memory of past states. This capability allows DCRNN to model changes in the data throughout the time series and predict future states accurately. 

\subsection{Spatiotemporal Data Preprocessing}
\label{sec:bkr-pre}
\begin{figure}[h]
 \centering
  \includegraphics[scale=0.70]{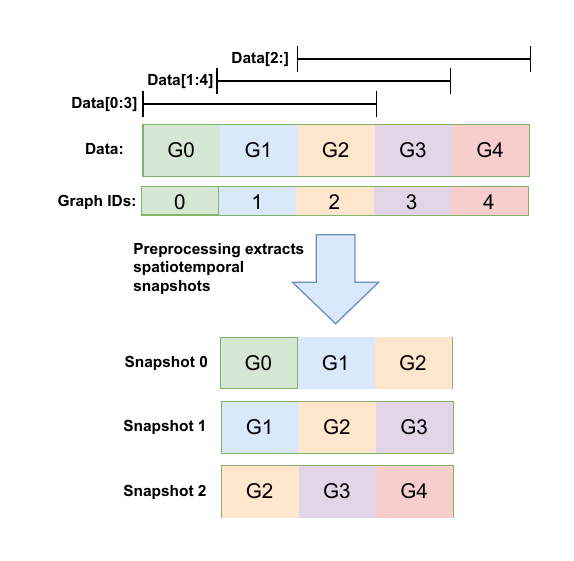}
\caption{Simplified workflow of spatiotemporal preprocessing with a $horizon$ of 3. G0 represents the state of the graph at time 0, G1 represents the state of the graph at time 1, and so on to time 4. During preprocessing, a sliding window of size  $horizon$ is applied to the data, extracting snapshots. After preprocessing, the data consists of a set of snapshots constructed from the original data. }
\label{fig:simplfiedPreprocessing}
\end{figure}

In order to convert data into a format compatible with ST-GNNs, the data must be restructured and standardized (the algorithm is shown in Algorithm~\ref{alg:preprocessing}). ST-GNNs rely on sequence-to-sequence data, where one sequence of spatiotemporal information is used as input to predict a subsequent sequence. Sliding window analysis (SWA)~\cite{Ockerman2023Accelerating} is used to segment the data into sequential, overlapping time slices (see \cref{fig:simplfiedPreprocessing}). SWA generates a corresponding target sequence ($y$) for each input sequence ($x$); $y$ is the sequence of a user-specified time period, which is referred to as the horizon, ahead of $x$. For example, if $x$ consists of the graphs of traffic speeds observed from 8:00 AM to 9:00 AM and the model is set to predict 60 minutes into the future, then $y$ would correspond to the ground truth graphs of traffic speeds from 9:00 AM to 10:00 AM. After this mapping is generated, the window is moved forward in time, and the process is repeated until the window has traversed the entire dataset. Subsequently, the data is normalized based on the training set, requiring the calculation of the mean and standard deviation. Normalization ensures that each node contributes equally to the model’s predictions, preventing those with larger values from dominating the learning process~\cite{Singh2020Investigating, Huang2020Normalization, Wan2019Influence}. 

The preprocessing algorithm described in algorithm~\ref{alg:preprocessing} has been applied to train many state-of-the-art ST-GNN models~\cite{zhu2020a3tgcnattentiontemporalgraph, zhao2019tgcn, li2021dynamic, Zhu2021_AST, roy2021_sst, Silu2023_STGC, liu2024_spatial}. Although effective for small datasets, the preprocessing algorithm and its common open-source implementations~\cite{li2021dynamic, rozemberczki2021_pytorch} are not designed with a focus on memory efficiency. \Cref{fig:simplfiedPreprocessing} shows a simplified visual illustration of spatiotemporal preprocessing, which extracts overlapping snapshots from the original data (the top half of \cref{fig:simplfiedPreprocessing}) and stores the collection of temporal snapshots in a list-based data structure (the bottom half of \cref{fig:simplfiedPreprocessing}). This process introduces a high degree of data redundancy and causes significant memory growth that precludes training on larger datasets without graph partitioning. Prior to this work, training a model on the full PeMS dataset, which grows to 419.46~GB during preprocessing, was intractable, causing open-source preprocessing implementations~\cite{li2018dcrnn_traffic, rozemberczki2021_pytorch} to crash even on a supercomputer node with 512~GB of RAM. For each dataset utilized in this work, the sizes before and after preprocessing, as well as key dataset characteristics, are shown in ~\cref{tab:bkrnd-datasets}.

\begin{algorithm}[!t]
\begin{algorithmic}[1]
\REQUIRE \texttt{integer}: horizon \COMMENT{How far we are predicting in the future}
\STATE file\_data = load(file\_path)
\STATE data = file\_data.get\_node\_data()
\STATE 
\STATE x,y = [], []
\STATE \# Let window be a 3D tuple defining the window shape
\FOR{each window in $data$}
    \STATE x.append(data[window])
    \STATE y.append(data[window + horizon])
\ENDFOR
\STATE 
\STATE \# Stack the spatiotemporal snapshots
\STATE x = stack(x, axis=0)
\STATE y = stack(y, axis=0)
\STATE 
\STATE \# Standardize the data 
\STATE x\_train = x[:round(len(x) * 0.70)]
\STATE $\mu$ = mean(x\_train)
\STATE $\sigma$ = std\_dev(x\_train)
\STATE x = (x - $\mu$) / $\sigma$  
\STATE y = (y - $\mu$) / $\sigma$
\RETURN  \texttt{x, y}
\end{algorithmic}
\caption{General algorithm for preprocessing spatiotemporal data based on popular open-source tools~\cite{rozemberczki2021_pytorch,li2018dcrnn_traffic,li2021dynamic}. Note that different methods of standardization and train, validation, and test splits are possible. }
\label{alg:preprocessing}
\end{algorithm}

\section{Motivating Case Study}
\label{sec:caseStudy}

To guide the design of PGT-I, we first investigate the challenges associated with scaling the ST-GNN workflow (data preprocessing, standardization, and training) to larger datasets using existing tools. To do so, we utilize two datasets: PeMS-All-LA, which posed significant challenges due to its high memory footprint in prior work \cite{Mallick2020Graph}, and PeMS~\cite{PEMS_Dataset}, which, to the best of our knowledge, is the largest available real-world spatiotemporal dataset. PeMS-All-LA is a medium-sized traffic dataset that includes graphs with 2,716 nodes, making it significantly larger than the datasets typically used for benchmarking (see \cref{tab:bkrnd-datasets}). PeMS contains 11,126 nodes and grows to nearly 420~GB after preprocessing. To the best of our knowledge, an ST-GNN has not been successfully trained on the full PeMS dataset without graph partitioning because of memory restrictions~\cite{Mallick2020Graph}.

We evaluate training behavior using two implementations of DCRNN. To establish a baseline, we utilize the PyTorch DCRNN implementation introduced in~\citet{li2018dcrnn_traffic}. Additionally, we modify the existing PGT-DCRNN implementation to support batching and stepwise sequence-to-sequence prediction. PGT is a popular ~\cite{Mohammadiyeh2023AnalyzingAI, Verdone2024_Explainable, Lin2022_Conditional} open-source Python library designed to facilitate the development and application of ST-GNNs. PGT models offer speedups over many prior benchmark ST-GNN implementations by leveraging optimized PyG implementations of graph operations, message passing, and parallelism~\cite{fey2019fastgraphrepresentationlearning}. To enable sequence-to-sequence prediction, we extend the model to process the input sequence in a stepwise fashion, feeding it one temporal slice at a time. Specifically, our implementation maintains and updates a hidden state across time steps, producing an output at each step that ultimately forms a prediction sequence of equal length to the input. While PGT-DCRNN implements the diffusion convolution operations described in \citet{li2018dcrnn_traffic}, it is a lightweight variant that uses a single spatiotemporal diffusion convolution layer and does not replicate the full behavior of the original model (e.g., the RNN-based encoder-decoder structure). As such, accuracy may differ between the two implementations. Our case study focuses on runtime and memory consumption, as these are critical factors for scalability, and the proposed optimizations (see \cref{sec:pgti-ib}) are applicable to both the original DCRNN and the simplified PGT-DCRNN.

\begin{table*}[h]
\begin{tabular}{|l|c|c|c|}
\hline
\textbf{Model} & \textbf{Runtime (mins)} & \textbf{Max System Memory Usage (GB)} & \textbf{Max GPU Memory Usage (GB)} \\ \hline
DCRNN & 68.48  & 371.25/512 & 24.84/40 \\ \hline
PGT-DCRNN & 4.48 & 259.84/512 & 1.58/40 \\ \hline

\end{tabular}
\centering
\vspace{5pt}
\caption{Single-epoch performance comparison of DCRNN and PGT-DCRNN using PeMS-All-LA.}
\label{tab:pemsAllLASingleEpoch}
\end{table*}

\subsection{Setup}
\label{sec:cs-setup}
 Given that we observe consistent resource usage across epochs, we limit node-hour costs by profiling training for a single epoch. We record the epoch runtime and  capture system memory usage and GPU memory usage on a per-second basis using \texttt{psutil} and \texttt{pynvml}, respectively. We perform testing on the state-of-the-art Polaris supercomputer using a single compute node. Each compute node is equipped with a 2.8~GHz AMD EPYC Milan 7543P 32-core CPU, 512~GB of DDR4 RAM, and four NVIDIA A100 GPUs. The system is interconnected using HPE Slingshot 11 and employs a Dragonfly topology with adaptive routing. We integrate the standard open-source preprocessing implementation~\cite{li2018dcrnn_traffic, li2021dynamic} into both workflows; and in all cases we use the default training, validation, and test split~\cite{li2018dcrnn_traffic}: 70\% train, 10\% validation, and 20\% test. Whenever possible, training uses the hyperparameters recommended by prior work~\cite{Mallick2020Graph}; however, because of GPU memory restrictions in the PyTorch DCRNN implementation, we select a batch size of 32. Since inference is not the focus of this work, we exclude evaluation with the test set from our experiments.

\subsection{Results}
\label{sec:cs}

\begin{figure}[b]
 \centering
  \includegraphics[width=\columnwidth]{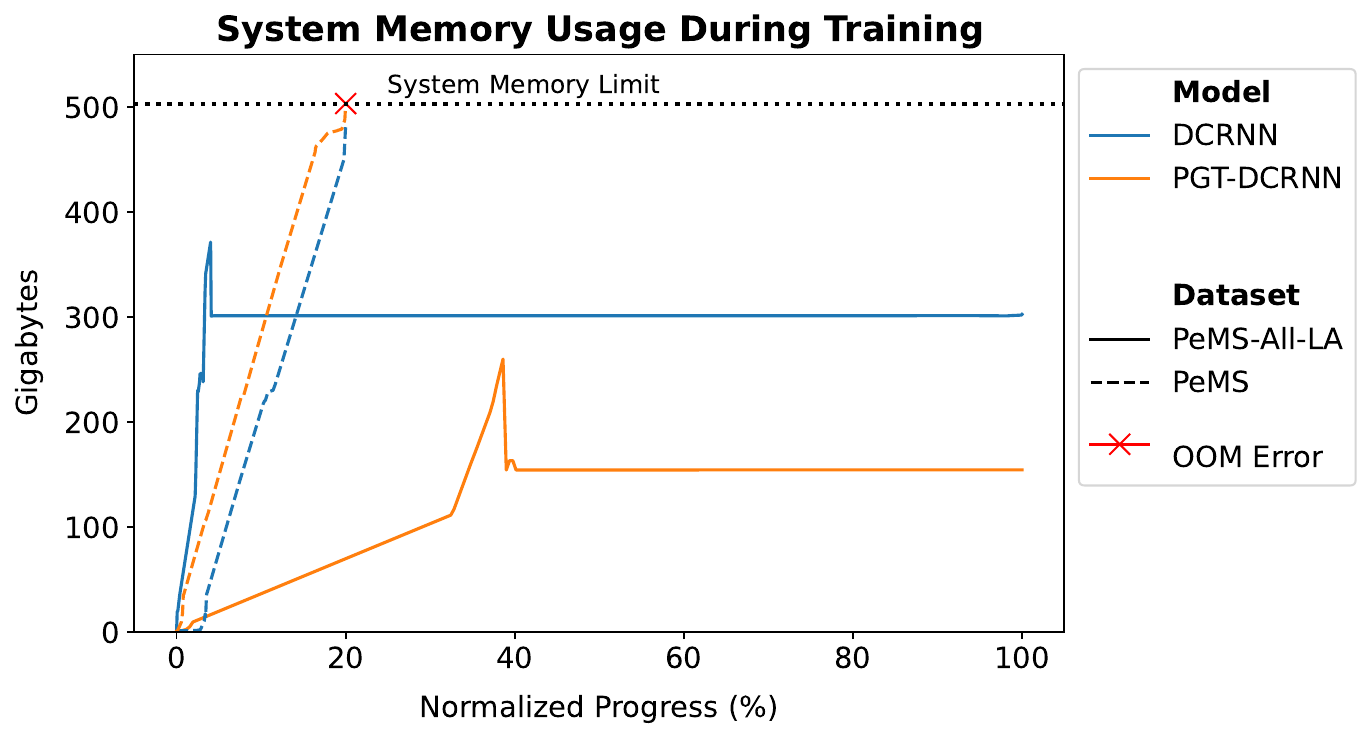}
\caption{Memory usage throughout training with PeMS-All-LA and PeMS using the original DCRNN model and the PGT-DCRNN model.}
\label{fig:caseStudyMemUsage}
\end{figure}

The PeMS-All-LA dataset presents significant memory challenges for both the DCRNN and PGT-DCRNN implementations. As shown in \cref{tab:pemsAllLASingleEpoch}, DCRNN exhibits high memory consumption, with a peak system memory usage of 371.25~GB and peak GPU memory usage of 24.84~GB. Although still suffering from high memory usage, the PGT-DCRNN model demonstrates substantial lower memory usage (see \cref{fig:caseStudyMemUsage}), reducing peak system memory usage to 259.84~GB and peak GPU memory usage to just 1.58~GB. The higher memory usage in the original DCRNN implementation stems from its custom dataloader, which stores extra copies of the dataset --- padded to align with the batch size --- in addition to the original data. PGT-DCRNN also reduces GPU memory usage and achieves a 15.30x reduction in runtime relative to the original DCRNN model, highlighting its potential as a lightweight foundation for integrating additional scalability techniques. 

Despite PGT-DCRNN's lower resource demands compared to DCRNN, neither implementation can scale to the full PeMS dataset because of system memory restrictions. As \cref{fig:caseStudyMemUsage} shows, both DCRNN implementations exceed the system memory limit and crash before beginning training. These results highlight that, unlike the typical challenge of excessive GPU memory usage in standard GNNs~\cite{Lee2024_addressing}, the primary scalability bottleneck for ST-GNNs lies in restrictive CPU memory consumption. 

\subsection{Analysis}
\begin{figure}[t]
 \centering
  \includegraphics[scale=0.45]{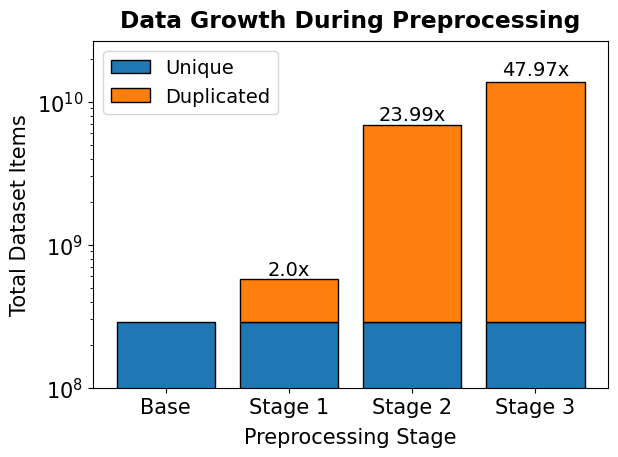}
\caption{Data growth when processing PeMS-All-LA. Stage 1 represents the added data from including time-of-day information as a transposed matrix. Stage 2 displays the data growth after applying sliding window analysis. Stage 3 shows the data growth by further dividing the data into heavily overlapping $x$ and $y$ train, validation, and test sets. }
\label{fig:preprocessing_data_growth}
\end{figure}

Motivated by DCRNN's high memory usage, we examined the ST-GNN workflow to identify the root causes. Further analysis revealed that the majority of the memory overhead stems from the representation of spatiotemporal graph snapshots produced by SWA. To generate the feature ($x$) and label ($y$) arrays, open-source implementations perform SWA on the source data, extracting an input sequence and label sequence at every valid placement of the window. As shown in \cref{fig:simplfiedPreprocessing}, preprocessing applies SWA to segment the data into time snapshots (stage 2 in \cref{fig:preprocessing_data_growth}). Each of the snapshots contains $horizon - 1$ redundant data values, where $horizon$ represents the number of time steps the model is predicting into the future. The data is then further duplicated by creating a $y$ time slice label for each corresponding $x$ time slice (stage 3 in \cref{fig:preprocessing_data_growth}). This process and its high degree of data growth are shown graphically for PeMS-All-LA in \cref{fig:preprocessing_data_growth}. The resulting data growth can be described analytically as follows: Given a spatiotemporal dataset whose original size is $size = entries \times nodes \times features$, its final size after preprocessing is described by \cref{eq:size}, where $entries$ is the total number of dataset items, $horizon$ is the number of time periods the model is predicting into the future, $nodes$ is the number of graph nodes in each graph, and $features$ is the number of data features (e.g., speed, day of week). 

\begin{align}
\text{size} &= 2 \big[(\text{entries} - (2 \times \text{horizon} - 1)) \notag \\
&\quad \times \text{horizon} \times \text{nodes} \times \text{features} \big] \label{eq:size}
\end{align}

As shown in \cref{fig:preprocessing_data_growth}, the current ST-GNN preprocessing workflow leads to significant data duplication, causing the majority of the postprocessed data to be redundant. The described data growth due to unnecessary data duplication is present in open-source ST-GNN tools~\cite{li2018dcrnn_traffic,rozemberczki2021_pytorch} and applies to many state-of-the-art models~\cite{zhao2019tgcn, li2021dynamic, Zhu2021_AST, roy2021_sst, Silu2023_STGC, liu2024_spatial, Huang2023_Spatio, rozemberczki2021_pytorch,ye2022_esg, zhu2020a3tgcnattentiontemporalgraph} in addition to DCRNN. To enable further scaling in the spatiotemporal family of models, we design methods tailored to address spatiotemporal data growth and integrate them into PGT-I.

\section{Design}
\Cref{sec:cs} demonstrates that even when using the full resources of a Polaris compute node, state-of-the-art tools such as PGT are insufficient for training models on large-scale spatiotemporal datasets. These challenges extend beyond DCRNN to the majority of ST-GNN models~\cite{zhao2019tgcn, li2021dynamic, Zhu2021_AST, roy2021_sst, Silu2023_STGC, liu2024_spatial, Huang2023_Spatio, rozemberczki2021_pytorch,ye2022_esg, zhu2020a3tgcnattentiontemporalgraph} because they similarly utilize sequence-to-sequence data and the standard ST-GNN preprocessing pipeline~\cite{li2018dcrnn_traffic,rozemberczki2021_pytorch} that results in significant memory growth. Our case study highlights the need for new tools that explicitly address spatiotemporal data growth and enable distributed, large-scale training. Thus, we propose techniques that decrease memory usage and enable distributed training without utilizing graph partitioning, which can negatively impact accuracy~\cite{Mallick2020Graph,Ji2024Local,hamilton2018inductiverepresentationlearninglarge}. Given its performance and flexibility, we select PGT as our base and propose an extension -- PyTorch Geometric Temporal-Index (PGT-I) -- designed to overcome current ST-GNN limitations and fully leverage HPC platforms to accelerate model training. PGT-I is a memory-efficient training framework that is fully integrated into the PGT code base and enables distributed training using Dask and Dask-DDP. Our implementation includes two novel techniques: 
\begin{itemize}
    \item Index-batching 
    \item Distributed-index-batching 
\end{itemize}

Index-batching introduces a new approach to spatiotemporal data management designed to reduce memory usage. To do so, it redesigns the preprocessing pipeline to eliminate data duplication and constructs the spatiotemporal snapshots at runtime using an array of graph IDs (see \cref{fig:simplfiedPreprocessing}). We provide both CPU and GPU implementations of index-batching. The latter, which we refer to as GPU-index-batching, extends index-batching to perform preprocessing and training entirely on the GPU. We combine our index techniques with distributed training, thereby enabling multi-node, multi-GPU computation within the ST-GNN workflow.

\subsection{Index-Batching}
\label{sec:pgti-ib}
\begin{figure}[h]
 \centering
  \includegraphics[width=\columnwidth]{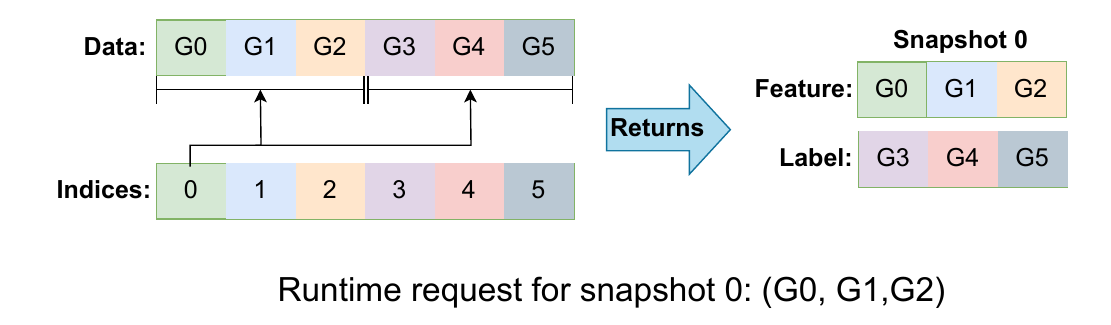}
\caption{Simplified example of constructing snapshots at runtime using index-batching where ``indices" store graph IDs. Given a horizon size (or window) of 3,  we construct the feature and label for snapshot 0 by extracting NumPy views from the data based on graph ID and horizon size.}
\label{fig:indexBatching}
\end{figure}

 Index-batching modifies ST-GNN preprocessing and training to significantly reduce memory usage. To do so, we take advantage of the unique properties of ST-GNN preprocessing. Given that all data contained in the generated $x$ feature array and the $y$ label array is already present in the original file, we observed that it is more efficient to store only a single copy of the original data (which corresponds to stage 2 in \cref{fig:preprocessing_data_growth}) and the graph IDs (see \cref{fig:indexBatching}) or indices corresponding to the data each $x$-$y$ mapping accesses. Furthermore, since the size of the time slices and the offset between $x$ and $y$ slices are constants defined by the $horizon$, it is unnecessary to store all the indices for each slice or maintain a separate set of indices for the $y$ array. For example, in \cref{fig:indexBatching}, which uses a $horizon$ of 3, if we store the graph ID that corresponds to the first entry in snapshot 0 (i.e., graph ID 0) in  $start$, we can reconstruct the snapshot and its label at runtime as NumPy views by accessing $data[start: start + horizon]$ and $data[start + horizon: start + (2 * horizon)]$. This practice removes the high degree of data duplication present in the standard preprocessing workflow by only storing an array of graph IDs roughly equal in size to the number of dataset entries ($ \text{entries} - 2 \times \text{horizon} - 1$) and a single copy of the original data. Additionally, by leveraging NumPy views to reference a given snapshot, we avoid copying data during batching. During training, batches are generated on demand by slicing into the standardized data using a set of indices and the $horizon$ value to reconstruct the appropriate snapshots.

Index-batching significantly improves memory efficiency compared with the existing ST-GNN workflow. As shown in \cref{eq:size}, the primary factors driving the increase in size during ST-GNN preprocessing are the $horizon$ size and the $2x$ multiplier due to maintaining separate sets of $x$ and $y$ data points. In contrast, index-batching's space needs are not increased by larger $horizon$ values, and it does not maintain a separate $y$ dataset in addition to the $x$ dataset. Index-batching's space requirements are represented by \cref{eq:index_size}, where 
$entries - (2 \times horizon - 1)$ represents the extra data due to including an array of indices. 

\begin{align}
\text{index\textunderscore batching\textunderscore size} &= \text{entries} \times \text{nodes} \times \text{features} \nonumber \\
&\quad + (\text{entries} - (2 \times \text{horizon} - 1)) 
\label{eq:index_size}
\end{align}

\noindent\textbf{GPU-Index-Batching:} To enable ST-GNN workflows to execute entirely on the GPU, we implement a GPU version of index-batching, which we refer to as GPU-index-batching. GPU-index-batching relies on the low memory footprint of index-batching and PyTorch tensor operations to handle preprocessing and training entirely on the GPU, eliminating mid-training CPU-to-GPU communication overhead. After reading the data from disk, the data is migrated from CPU memory to GPU memory. Once on the GPU, the workflow proceeds in the same manner as index-batching, simulating the effects of SWA without requiring explicit data duplication. 

In contrast to previous techniques that limit optimizations to GNN training~\cite{Swapnil2021_P3,guan2022dynagraph,wang2020deep,Zheng2020Dist,Zhou2023_DistTGL}, we extend our optimizations to preprocessing. Further, by transferring all data to GPU memory prior to the start of training, we eliminate the overhead of periodically moving data between the CPU and GPU during training, which is particularly relevant with larger datasets. While the GPU-index-batching method is not suitable for datasets that exceed GPU memory capacity, the current largest real-world ST-GNN dataset PeMS~\cite{PEMS_Dataset} is 8~GB in size, and the majority of ST-GNN benchmark datasets \cite{rozemberczki2021_pytorch,Mallick2020Graph} are less than 2.5~GB in size (see \cref{tab:bkrnd-datasets} for more examples). By eliminating the data growth typically associated with ST-GNN preprocessing, GPU-index-batching offers a practical and efficient technique for spatiotemporal datasets.

\subsection{Distributed-Index-Batching}

Distributed-index-batching combines DDP training and index-batching techniques to enable memory-efficient, distributed ST-GNN workflows. To enable DDP training, we utilize \texttt{dask-pytorch-ddp},~\footnote{\url{https://github.com/saturncloud/dask-pytorch-ddp}} a wrapper around PyTorch's DDP module that allows it to leverage \texttt{Dask.distributed}. We integrate Dask-DDP into PGT, following the standard DDP workflow present in popular open-source distributed tools~\cite{torch_ddp,wolf2020huggingfacestransformersstateoftheartnatural}. In order to mirror the baseline open-source DCRNN implementation~\cite{li2018dcrnn_traffic},\footnote{\url{https://github.com/liyaguang/DCRNN/blob/master/lib/utils.py\#L189}} the dataset is shuffled at the start of each epoch. After shuffling, the dataset is partitioned according to the number of workers, enabling each worker to independently perform forward and backward passes. During the backward pass, each worker computes the gradients of the loss with respect to the model parameters. These gradients are then averaged across all workers through an all-reduce operation, and the averaged gradient is used to update each local model’s parameters.

When using distributed-index-batching, each worker maintains its own in-memory copy of the dataset, allowing it to perform preprocessing and training entirely in local memory without requiring inter-worker communication (aside from DDP calls to \texttt{AllReduce} to average model gradients). This is possible due to the low memory footprint of index-batching; previously, even a single copy of PeMS would exceed a Polaris compute node's 512~GB of memory during preprocessing. Maintaining a local copy in each worker's memory enables communication-free global shuffling, that is, shuffling across all workers between epochs. At the start of each epoch, each worker samples a new data subset from its local memory without inter-worker communication. Global shuffling differs from local shuffling strategies, which keep the data each worker uses, referred to as its data partition, fixed across epochs. In local shuffling, data is shuffled only within a worker’s own partition, without any mixing between partitions. Previous work demonstrated that local shuffling can result in slower convergence~\cite{Nguyen2022WhyGlobally, meng2017convergenceanalysisdistributedstochastic} and reduced final accuracy~\cite{meng2017convergenceanalysisdistributedstochastic}. Therefore, we elect to utilize global shuffling. By default, distributed-index-batching utilizes GPU-index-batching at the worker level; however, similar to our single-GPU techniques, we also support a CPU-based alternative.

\section{Evaluation}
\label{sec:eval}
PGT-I has the potential to retain the ease of use of PGT, reduce memory usage in a single-GPU setting, and provide multi-GPU and multi-node scalability through specialized distributed training techniques. To evaluate our techniques, we aim to answer three key questions:
\begin{enumerate}
    \item What is the impact of index-batching on runtime, memory usage, and accuracy in a single-GPU setting?
    \item How does GPU-index-batching affect memory usage and runtime relative to standard index-batching in a single-GPU setting?
    \item How does distributed-index-batching scale with respect to runtime and accuracy as the number of GPUs increases, particularly relative to standard DDP techniques?  
\end{enumerate}

To test PGT-I's impact on accuracy, runtime, and memory usage in a single-GPU setting, we perform training using three benchmark datasets: PeMS-Bay~\cite{li2018dcrnn_traffic}, Windmill-Large~\cite{rozemberczki2021_pytorch}, and Chickenpox-Hungary~\cite{rozemberczki2021chickenpoxcaseshungarybenchmark}. These datasets represent several common ST-GNN benchmarks and span multiple domains, making them an ideal baseline for testing PGT-I's broader applicability. We train with batch sizes of 64, 64, and 4, respectively. The first two batch sizes are selected based on prior work's hyperparameter tuning~\cite{li2018dcrnn_traffic}, while Chickenpox-Hungary's batch size is selected based on its limited 522 entries. To enhance performance and encourage generalizability across datasets, we adopt established hyperparameters \cite{Mallick2020Graph} and the default PyTorch Adam optimizer. We compare the performance of PGT-DCRNN with standard batching to its performance with index-batching. Each training run consists of 100 epochs. We measure runtime, accuracy, and memory usage and average the results over 10 runs on Polaris (see \cref{sec:cs-setup} for hardware specifications) to increase robustness. Additionally, to study the effect of dataset scale on memory savings, we compare PGT's and PGT-I's memory usage with the full PeMS dataset.

To investigate the performance of single-GPU index-batching and GPU-index-batching, we conduct model training with the full PeMS dataset on Polaris. The models are trained for 30 epochs, utilizing the Adam optimizer and model hyperparameters found in past work~\cite{Mallick2020Graph,li2018dcrnn_traffic}. To evaluate distributed performance, we conduct a scaling study with the PeMS dataset, collecting data on runtime, accuracy, and memory usage across 30 epochs. As is standard practice in ML scaling studies~\cite{kaplan2020scalinglawsneurallanguage,rajbhandari2020zeromemoryoptimizationstraining,krizhevsky2014weirdtrickparallelizingconvolutional}, the dataset remains fixed, and the global batch size (i.e., the batch size per worker $\times$ the number of workers) increases as we increase the number of GPUs. We perform distributed training using 4, 8, 16, 64, and 128 GPUs (corresponding to 1, 2, 4, 8, 16, and 32 compute nodes, respectively, on Polaris). 

\begin{table}[b]
\scalebox{0.75}{
\begin{tabular}{|l|c|c|c|}
\hline

 & \textbf{Runtime (s)} & \textbf{MAE} & \textbf{Max Memory Usage (~MB)} \\ \hline
Base-Chickenpox & 188 $\pm$ 5.15  & 0.6061 $\pm$ 0.0011 & 1093 $\pm$ 2.24 \\ \hline
Index-Chickenpox & 192 $\pm$ 3.2  & 0.6061 $\pm$ 0.0010 & 1089 $\pm$ 4.09 \\ \hline

Base-Windmill & 2323 $\pm$ 10.86  & 0.1707 $\pm$ 0.0303 & 2455 $\pm$ 72.93 \\ \hline
Index-Windmill & 2339 $\pm$ 18.78  & 0.1606 $\pm$ 0.0231 & 1304 $\pm$ 39.99 \\ \hline

Base-PeMS-Bay & 3731 $\pm$ 36.32 & 1.8923 $\pm$ 0.0056 & 4497 $\pm$ 56.67 \\ \hline
Index-PeMS-Bay & 3735 $\pm$ 31.41  & 1.8892 $\pm$ 0.0055 & 1335 $\pm$ 62.13 \\ \hline

\end{tabular}}
\centering
\vspace{5pt}
\caption{Performance comparison of base PGT-DCRNN and PGT-DCRNN utilizing index-batching. Runtime and memory statistics are the average of 10 experiments performed on Polaris.}
\label{tab:comparison}
\end{table}
Given that existing distributed graph training libraries (e.g.,  P3~\cite{Swapnil2021_P3}, DGL~\cite{wang2020deep}, and DIST-DGL~\cite{Zheng2020Dist}) are incompatible with spatiotemporal data, we compare distributed-index-batching with its single-GPU PGT counterpart and a standard DDP workflow, which we refer to as baseline DDP or simply DDP. As described in \cref{sec:caseStudy}, PGT represents the state of the art in spatiotemporal training and integrates a range of performance optimizations, making it an ideal single-GPU baseline. Additionally, we compare performance relative to DDP, implemented using PGT for computational efficiency and \texttt{Dask.distributed} to distribute data across the workers during both preprocessing and training; similar to many large-scale DDP frameworks~\cite{Zheng2020Dist,elgabli2020gadmmfastcommunicationefficient,Shao2024Distributed}, this method communicates data on demand across workers. To reduce the cost of communication and ensure a fair comparison, we tested multiple DDP implementations and a variety of Dask configurations (e.g., data transfer block size, threads per worker, spill-to-disk threshold), selecting the best-performing configuration. We found that issuing Dask communication requests for a batch of data rather than individual data items within a batch, which would align with the standard PyTorch sampler's practice, significantly reduced communication overhead; hence, we incorporated this optimization into our baseline DDP approach. In order to maintain parity with our distributed-index-batching approach and avoid the decreased accuracy associated with local shuffling~\cite{meng2017convergenceanalysisdistributedstochastic}, DDP performs global shuffling between epochs.

 \begin{figure*}[h]
 \centering
  \includegraphics[scale=0.35]{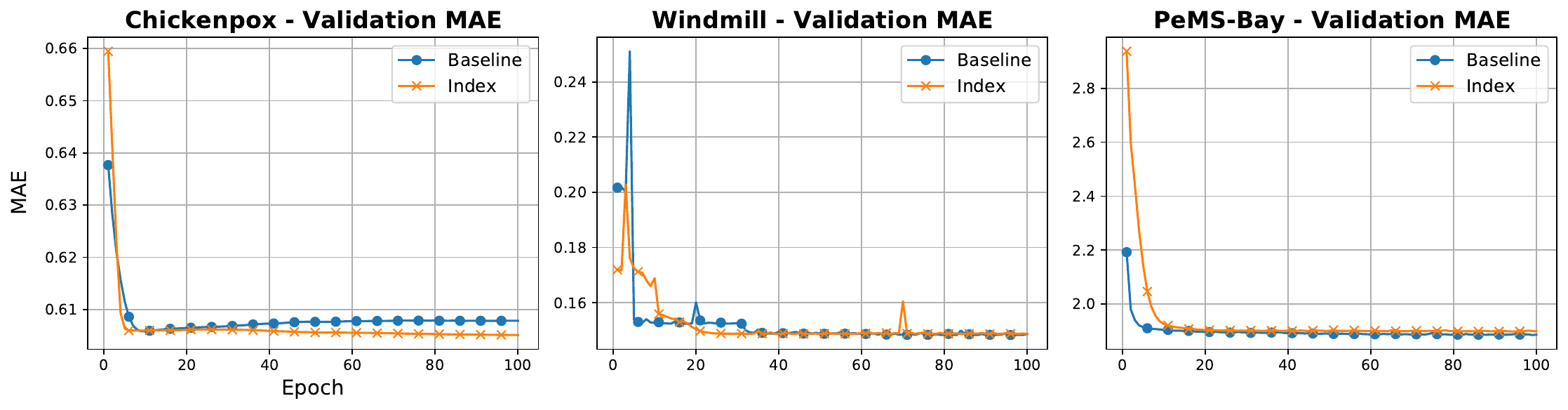}
\centering
\caption{ Single-GPU validation accuracy for each dataset during training. Notably, index-batching provides  accuracy and convergence speed comparable to PGT with standard batching techniques.}
\label{fig:acc_single_gpu}
\end{figure*}

 \begin{figure}[b]
 \centering
  \includegraphics[scale=0.45]{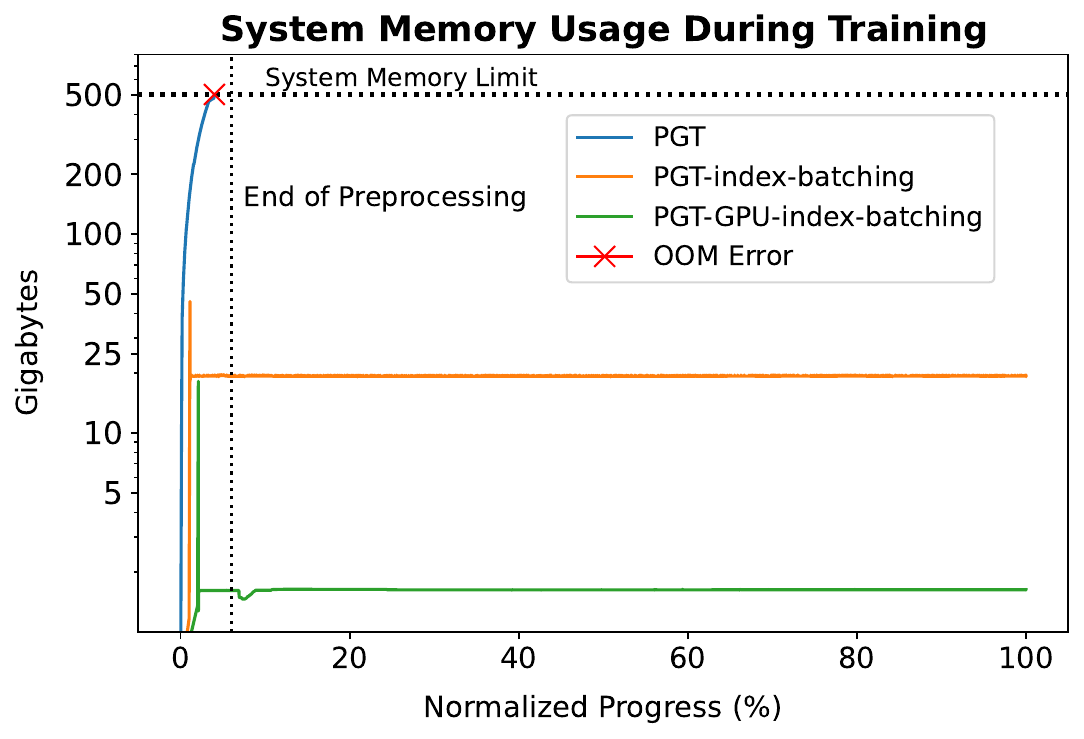}
\caption{Single-GPU memory usage with PeMS with and without index-batching. While memory usage is presented as measured, the positions of index-batching and GPU-index-batching's measurements have been shifted right by 1\% and 2\%, respectively, along the x-axis for visual clarity. }
\label{fig:pems_single_gpu}
\end{figure}

\subsection{Effect of Single-GPU Index-Batching on Accuracy, Runtime, and Memory Usage}
 \looseness-1As shown in \cref{tab:comparison}, index-batching reduces memory usage while demonstrating less than 1\% absolute difference in overall runtime. The memory reduction is proportional to the size of the dataset and the horizon. Thus, with smaller PGT datasets such as Chickenpox-Hungary, which grows to only 643~KB after preprocessing, the memory reduction is minimal. For more modest datasets such as Windmill-Large and PeMS-Bay, 
 index-batching achieves clear memory overhead reductions of 46.88\% and 70.31\%, respectively. The effect is even more pronounced with the full PeMS dataset; \cref{fig:pems_single_gpu} shows the memory usage of the baseline (standard ST-GNN batching techniques), index-batching, and GPU-index-batching implementations. The standard PGT-DCRNN workflow crashes during preprocessing because it exceeds the system memory limit of 512~GB. In contrast, PGT-DCRNN uses a maximum of 45.75~GB of memory with index-batching, enabling training on large datasets even on commodity devices.

\Cref{fig:acc_single_gpu} displays the training and validation accuracy of PGT for each dataset based on a single test iteration, as we observe minimal variation across runs (see \cref{tab:comparison}). While the inherent randomness of ML training affects the convergence process, as the training progresses, each batching technique ultimately exhibits similar convergence times and negligible differences in optimal MAE values (see \cref{tab:comparison} for exact numbers). This behavior is expected; index-batching feeds the same spatiotemporal snapshots to the model as standard ST-GNN batching, providing identical accuracy while reducing memory overhead.

\subsection{Benefits of GPU-Index-Batching Relative to Index-Batching}
As shown in \cref{tab:gpu_index_metrics}, GPU-index-batching accelerates data preprocessing and model training over index-batching on the PeMS dataset, reducing runtime by 12.87\%. Both implementations complete preprocessing in under 30 seconds, with index-batching requiring 26.05 seconds and GPU-index-batching 19.05 seconds. Since preprocessing accounts for only a small fraction of the total training time, we did not prioritize optimizing it for GPU execution. GPU-index-batching's speedup is due to the elimination of CPU-to-GPU data transfers during training. As a result of storing the dataset in GPU memory, however, GPU-index-batching uses significantly more GPU memory than does index-batching, requiring 18.60~GB relative to index-batching's 5.50~GB. This is balanced by reducing CPU memory usage by 60.30\%. As shown in \cref{fig:pems_single_gpu}, index-batching causes a spike in CPU memory usage to approximately 46~GB. This initial spike occurs during preprocessing, and by performing preprocessing on the GPU, GPU-index-batching reduces the spike and maintains a lower overall memory footprint.

\begin{table}[t]
\scalebox{0.9}{
\begin{tabular}{|c|c|c|c|}
\hline
\textbf{Implementation}& \textbf{Runtime} & \textbf{CPU Memory} & \textbf{GPU Memory}\\
\hline
Index-batching & 333.58 min & 45.84~GB &  5.50~GB\\
\hline
GPU-index-batching & 290.65 min & 18.20~GB &  18.60~GB \\
\hline
\end{tabular}}
\centering
\vspace{5pt}
\caption{Single-GPU PeMS training performance metrics.}
\label{tab:gpu_index_metrics}
\vspace{-0.20in}
\end{table}

\subsection{Distributed-Index-Batching Scaling Study}
This section presents the results of our scaling study, beginning with an evaluation of our solution's strong scaling relative to its single-GPU performance. We then compare the performance of distributed-index-batching with that of DDP and conclude with a discussion of the impact of increasing parallelism on model accuracy.

\subsubsection{Scaling}
\label{sec:scaling}
Distributed-index-batching demonstrates favorable scaling relative to its single-GPU implementation, reducing runtime by up to 79.41x with 128 GPUs when including preprocessing and up to 115.49x when considering training time alone. As anticipated, the static preprocessing cost and overhead introduced by DDP (e.g., syncing gradients, \texttt{AllReduce} operations to calculate validation accuracy) hinder linear scaling, especially as overall runtime decreases. Our technique achieves near-linear training scaling with 4, 8, 16, and 32 GPUs but falls short with 64 and 128 GPUs since fixed costs constitute a larger proportion of the total runtime.

\begin{figure}[t]
 \centering
  \includegraphics[width=\columnwidth]{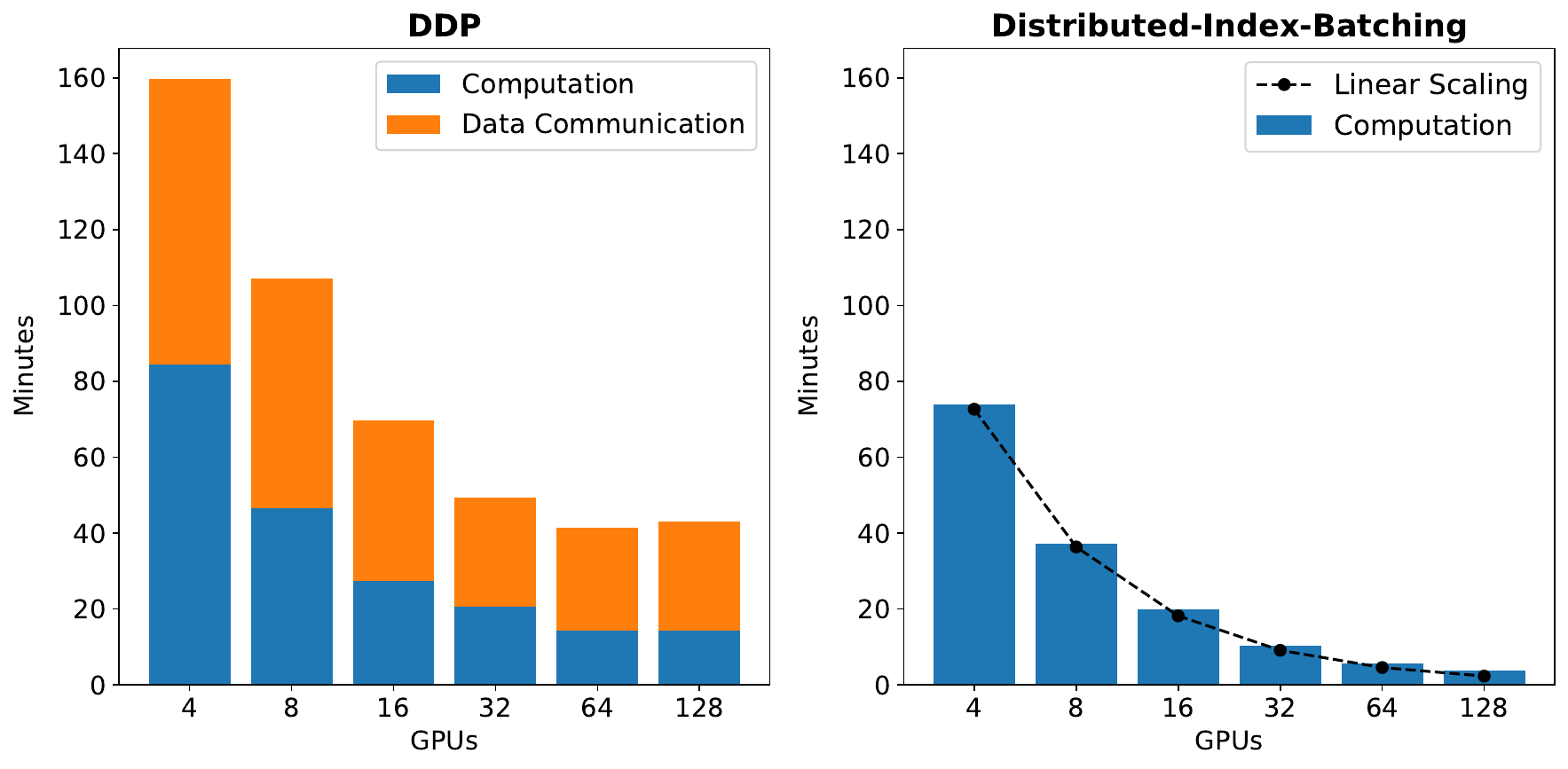}
\caption{Scaling study runtime results. The x-axis shows the number of GPUs/workers, while the y-axis shows the total runtime in minutes. By eliminating the need for inter-worker data communication, distributed-index-batching significantly outperforms DDP.}
\label{fig:eval-ddpPre}
\end{figure}

PGT-I's runtime efficiency significantly lowers the barrier to performing training with the PeMS dataset, which was previously intractable even on a state-of-the-art compute node. We note that preprocessing time with distributed-index-batching fluctuates because of the need to perform I/O over a shared parallel file system and Dask setup time. This is evident in the abnormally high preprocessing time of approximately 35-40 seconds with 16 and 32 GPUs in contrast to the 10–20 seconds observed in the majority of index-batching experiments. In follow-up experiments with 4 GPUs, we observed preprocessing times ranging from 11 seconds to 32 seconds, corresponding to measured fluctuations in I/O time rather than data preprocessing. These I/O fluctuations exist regardless of the number of workers. Nevertheless, despite I/O variability, PGT-I maintains fast and scalable training performance, making large-scale spatiotemporal modeling more accessible.

\subsubsection{Comparison with DDP}
Distributed-index-batching significantly outperforms the DDP approach, reducing overall runtime by 2.16x and 11.78x with 4 and 128 GPUs, respectively. This speedup arises from the elimination of inter-node communication and periodic CPU-to-GPU memory transfers, since workers can process data in local memory and utilize a single consolidated CPU-to-GPU memory transfer. As noted in \cref{sec:scaling}, because each worker performs preprocessing in local memory, distributed-index-batching's preprocessing time does not scale with the number of GPUs.

Distributed-index-batching demonstrates significantly better scaling than does DDP. As the number of workers increases, communication overhead limits DDP's scaling. Furthermore, DDP's preprocessing time remains relatively stable, reaching a maximum of 305 seconds with 128 workers, despite distributing computation across more workers. The increase in preprocessing time with 128 workers is due to the overhead of distributing data across a larger number of workers, which outweighs the benefits of increased preprocessing parallelization with PeMS. However, DDP maintains a smaller memory footprint than distributed-index-batching when using 32, 64, and 128 GPUs. When using 32 workers, for example, DDP reduces its maximum memory footprint to 53.3~GB, falling below distributed-index-batching's footprint of 90.18~GB.

\begin{figure}[t]
 \centering
  \includegraphics[width=\columnwidth]{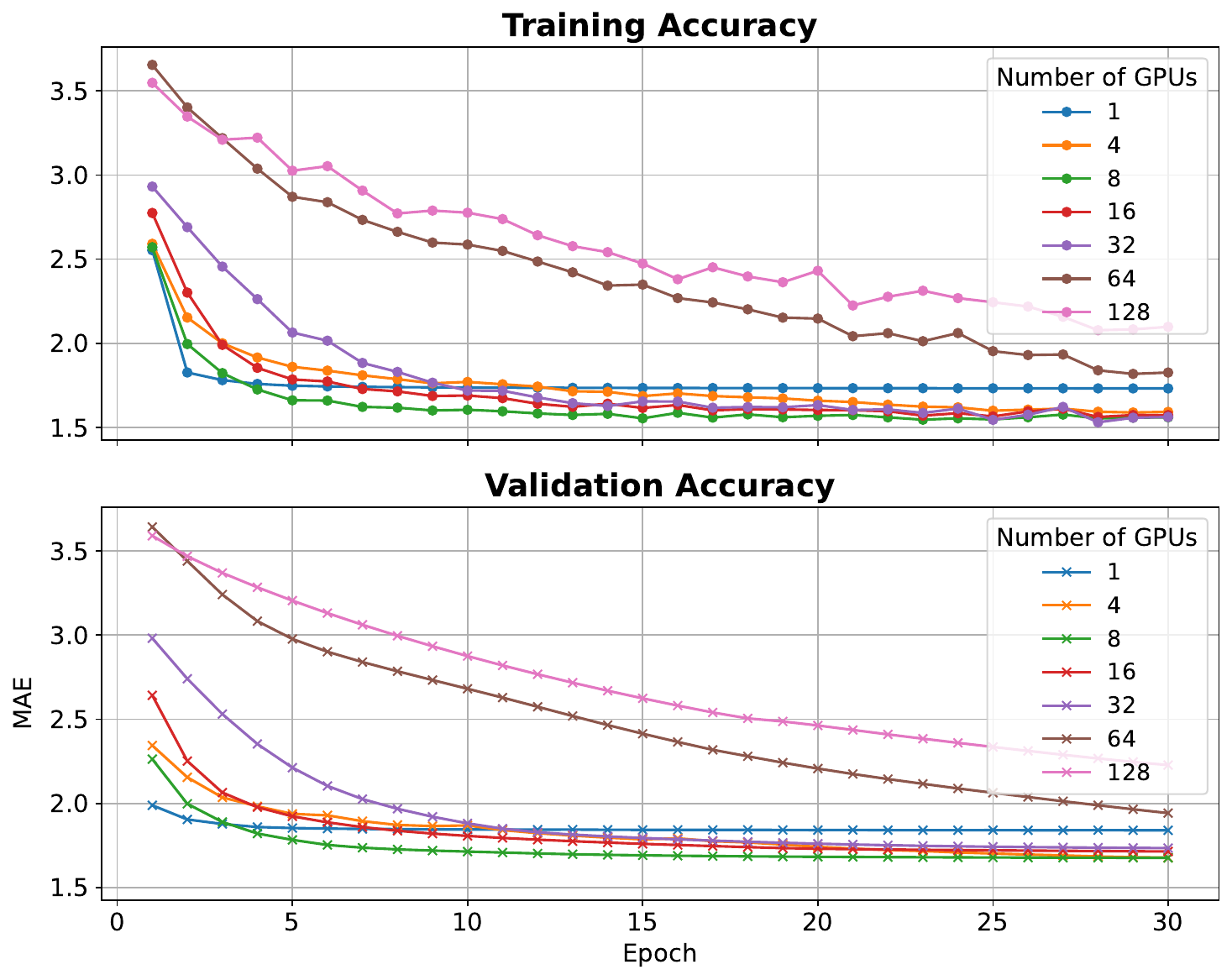}
\caption{Training and validation MAE on the PeMS dataset as the number of GPUs increases.}
\label{fig:eval-acc}
\end{figure}

\subsubsection{Accuracy with Increasing GPUs}
\label{sec:moreGPUsaccuracy}
\Cref{fig:eval-acc} presents the training and validation MAE as the number of GPUs increases. As the number of GPUs used for DDP increases, the optimal training and validation MAE also increases, with a value of 1.66 with 1 GPU and a value of 2.23 with 128 GPUs. This effect has been established in prior literature~\cite{chaubard2024gradientaveragingparalleloptimization,you2017largebatchtrainingconvolutional,goyal2018accuratelargeminibatchsgd}; averaging gradients over an increasing number of distributed mini-batches, as well as an increasingly large global batch size, can alter convergence dynamics and may result in diminished learning. This presents a trade-off between final accuracy and computational efficiency. However, the decline in accuracy with more GPUs may be offset by significantly improved runtime. Follow-up testing indicated that the majority of the MAE increase was due to the increased global batch size rather than DDP effects, and techniques such as learning rate scaling \cite{you2017largebatchtrainingconvolutional} reduced the increase in MAE with larger global batch sizes.  

\subsection{Scaling to Larger-than-Memory Datasets}
To enable training with larger-than-memory datasets, we design a version of distributed-index-batching that supports distributed preprocessing and partitions data across all workers. We refer to this approach as generalized-distributed-index-batching. To improve on our baseline DDP implementation, we seek to reduce communication overhead by utilizing a fixed data partition and replacing global shuffling with batch-level shuffling. That is, rather than shuffle individual samples, the order of the batches is shuffled (while the data inside the batch remains fixed) within each worker's data partition. This approach improves memory locality and reduces the number of separate communications required to load a batch into local memory. To assess the impact on accuracy of local batch shuffling, we compare accuracy with global shuffling with PeMS-Bay using 4, 8, and 16 GPUs and the same experimental settings described in \cref{sec:eval}. The results are shown in \cref{tab:shufflingVal}. Notably, local batch-level shuffling obtains accuracy similar to that of global shuffling.

\begin{table}[t]
\centering
\begin{tabular}{|l|c|c|c|}
\hline
\textbf{Implementation} & \textbf{4 GPUs} & \textbf{8 GPUs}  & \textbf{16 GPUs} \\
\hline
Global Shuffling &  1.932 & 2.008 & 2.149 \\
\hline
Local Batch Shuffling & 1.913 & 1.868 & 1.833 \\
\hline  

\end{tabular}
\centering
\vspace{5pt}
\centering
\caption{Optimal Validation MAE with global shuffling vs local batch shuffling using PeMS-Bay.}
\label{tab:shufflingVal}
\end{table}

To evaluate the runtime performance of our technique, we utilize the larger PeMS dataset, performing a single epoch of training using 4, 8, 16, 32, 64, and 128 GPUs. We compare with DDP modified to include the changes described above. Notably, while index-batching reduces distributed preprocessing time relative to DDP across all configurations, preprocessing constitutes a small portion of the overall runtime in a full training workflow. Therefore, we omit further analysis of preprocessing time.

\begin{figure}[b]
 \centering
  \includegraphics[width=\columnwidth]{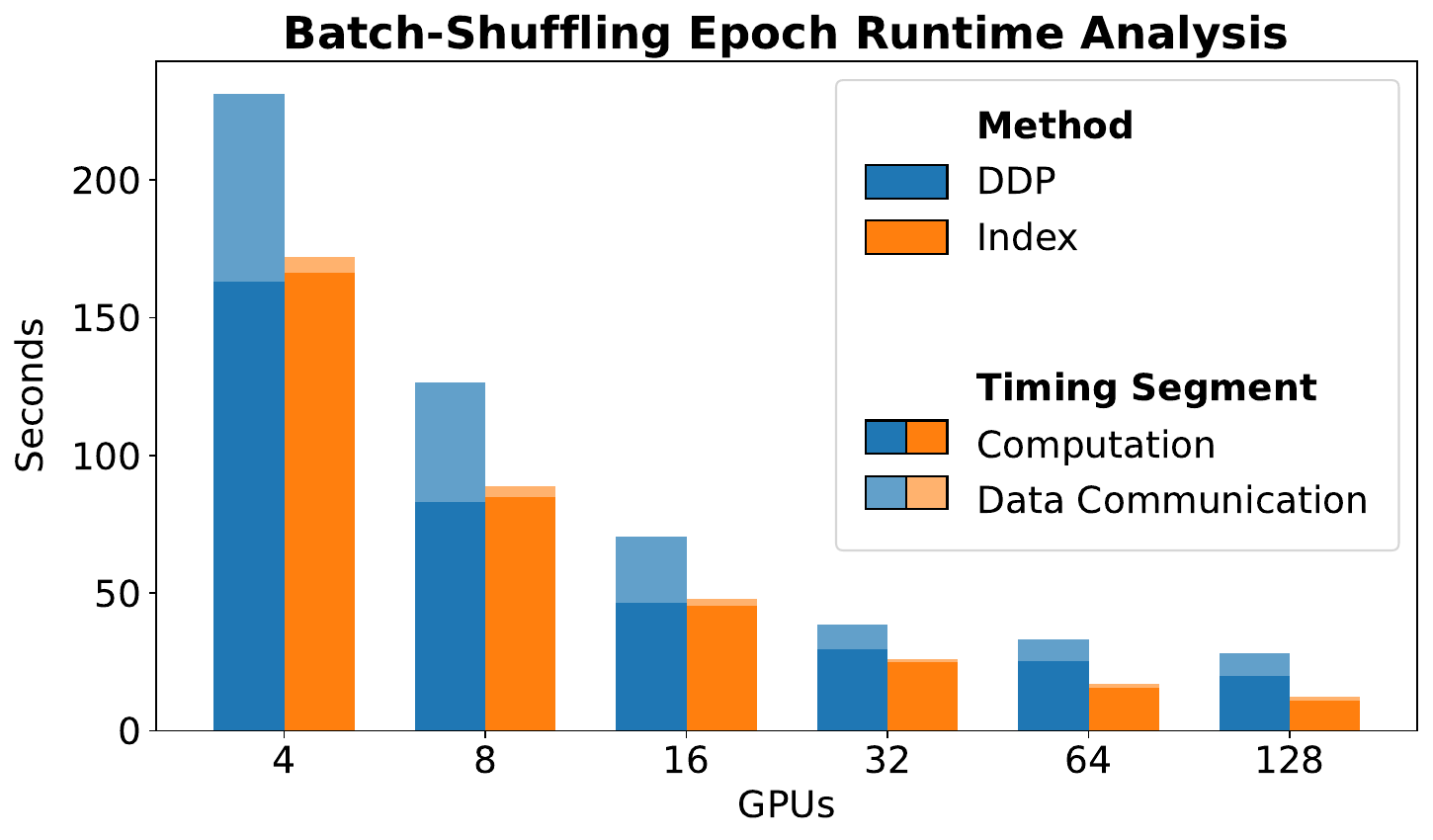}
\caption{Batch-shuffling versions of generalized-distributed-index-batching and baseline DDP. Generalized-distributed-index-batching significantly lowers communication cost (as shown by the lighter colored portion of each bar) by decreasing overall data volume. }
\label{fig:optTimes}

\end{figure}

Generalized-distributed-index-batching outperforms the baseline epoch time by up to 2.28x (see \cref{fig:optTimes}). While the baseline's epoch time improves from 303 seconds using 4 GPUs to 231 seconds, it still suffers from high overhead due to the large volume of communicated data. Generalized-distributed-index-batching also significantly reduces memory usage, using only 53.28~GB of memory with four workers compared with the baseline's 479.66~GB. Notably, the index-batching implementation's memory footprint with only 4 GPUs is comparable to the footprint baseline DDP achieved with 32 GPUs (53.30~GB), demonstrating substantial memory savings even in a single-node setting.

\subsection{Broader Applicability}
To demonstrate index-batching’s generalizability to sequence-to-sequence models, we perform additional single-GPU experiments with A3T-GCN \cite{zhu2020a3tgcnattentiontemporalgraph} and the Metr-LA \cite{PEMS_Dataset} dataset. Additionally, we perform a distributed-index-batching scaling study with ST-LLM \cite{liu2024_spatial} and the PeMS-Bay dataset. A3T-GCN is a spatiotemporal forecasting model that integrates graph convolutions with gated recurrent units and a global attention \cite{vaswani2023attentionneed} mechanism to achieve high accuracy. We integrate index-batching into A3T-GCN's PGT implementation, \footnote{\url{https://github.com/benedekrozemberczki/pytorch_geometric_temporal/blob/master/examples/recurrent/a3tgcn2_example.py}} utilizing the provided hyperparameters and training for 30 epochs. ST-LLM utilizes a state-of-the-art approach that encodes spatial-temporal context into token embeddings that are then processed by GPT2 \cite{Radford2019LanguageMA}. ST-LLM is not currently integrated into PGT, and  we therefore use its open-source implementation.\footnote{\url{https://github.com/ChenxiLiu-HNU/ST-LLM}} All tests are performed on the Polaris supercomputer, which is described in \cref{sec:cs-setup}, and we reuse the experimental settings described in \cref{sec:eval}. To reduce node-hour cost, we limit the distributed-index-batching tests to 4, 8, 16, and 32 GPUs.

\begin{table}[t]
\scalebox{0.9}{
\begin{tabular}{|c|c|c|c|}
\hline
\textbf{Implementation}& \textbf{Runtime} & \textbf{CPU Memory} & \textbf{Test MSE} \\
\hline
Baseline & 1041.95 (s) & 2426.26~MB &  0.5436\\
\hline
Index-batching & 1050.80 (s) & 1232.62~MB &  0.5427\\

\hline
\end{tabular}}
\centering
\vspace{5pt}
\caption{Single-GPU A3T-GCN performance metrics.}
\label{tab:a3_gpu_index_metrics}
\vspace{-0.20in}
\end{table}

\begin{figure}[t]
 \centering
  \includegraphics[width=\columnwidth]{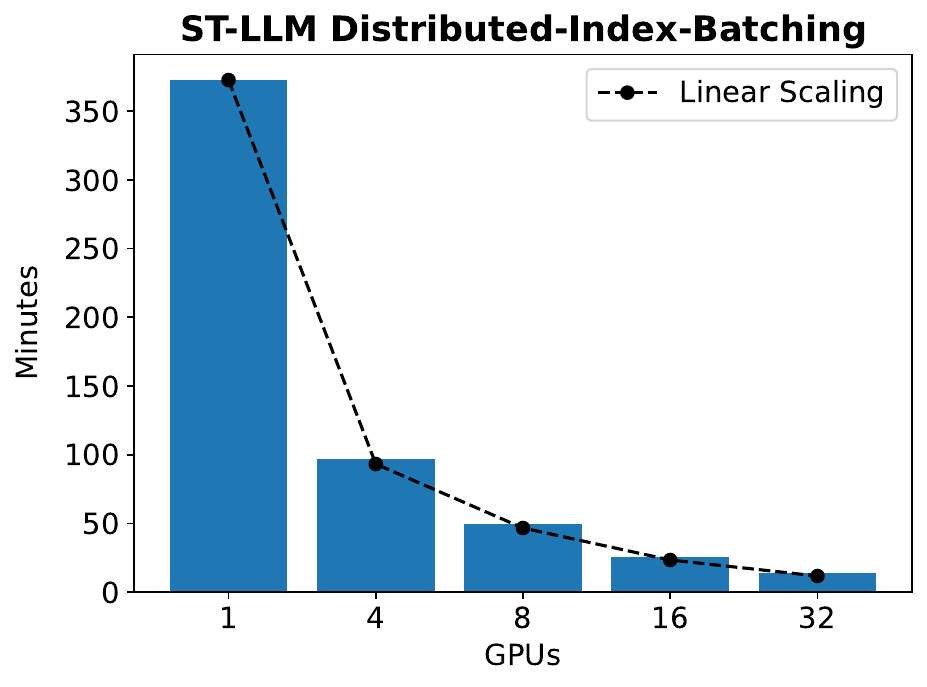}
\caption{Scaling study runtime results with ST-LLM. The x-axis shows the number of GPUs/workers, while the y-axis shows the total runtime in minutes.}
\label{fig:eval-ddpSTLMM}
\end{figure}

Single-GPU tests with A3T-GCN reinforce that index-batching reduces memory usage with negligible impact on runtime or accuracy, achieving a 49.20\% reduction in memory usage (see \cref{tab:a3_gpu_index_metrics}). Additionally, distributed-index-batching reduces training time by 3.92x with 4 GPUs and 30.01x with 32 GPUs compared with single-GPU index-batching. Due in part to the small size of PeMS-Bay, the overall workflow demonstrates near-linear scaling (see \cref{fig:eval-ddpSTLMM}); and preprocessing constitutes a minimal portion of overall runtime, requiring at most 1.35 seconds. These results not only highlight the advantages of index-batching but also demonstrate its broader applicability to sequence-to-sequence models.

\subsection{Discussion and Summary of Key Results}

Index-batching enables ST-GNNs to train on formerly intractable datasets such as PeMS using a single GPU. Additionally, distributed-index-batching outperforms a baseline DDP implementation, demonstrating near-linear scaling. By leveraging the memory savings provided by index-batching, distributed-index-batching is able to maintain a local copy of the dataset in each worker's memory and eliminate the communication required for global shuffling. However, this optimization is not applicable to datasets larger than memory, where data must be distributed across multiple nodes. To the best of our knowledge, such spatiotemporal datasets do not currently exist; however, in the future, larger-than-memory datasets likely will be made available. Thus, we include an extension to distributed-index-batching that integrates distributed data management techniques to ensure future scalability. 

\section{Related Work}
A significant amount of effort has also been dedicated to efficient distributed GNN training~\cite{Besta2024Parallel,Zheng2020Dist}. \citet{Swapnil2021_P3}, \citet{wang2020deep}, and \citet{Zheng2020Dist} present distributed GNN training libraries P3, DGL, and DIST-DGL. However, these libraries are not compatible with spatiotemporal data. As such, they are unable to address the need for distributed spatiotemporal training and are incompatible as benchmarks for comparison with our techniques. Another relevant distributed graph training library is DistTGL~\cite{Zhou2023_DistTGL}. DistTGL is a distributed training framework designed specifically for training temporal graph neural networks. However, DistTGL lacks support for spatiotemporal coupling, where graph-based spatial dependencies and sequential temporal modeling must be tightly integrated. 

To the best of our knowledge, DynaGraph~\cite{guan2022dynagraph} is the only proposed framework designed specifically for distributed ST-GNN training. However, DynaGraph utilizes graph partitioning, introducing overhead due to communication \cite{Chen2019_PowerLyra} and potentially reducing accuracy~\cite{Mallick2020Graph,Ji2024Local,hamilton2018inductiverepresentationlearninglarge}. To reduce memory requirements, \citet{guan2022dynagraph} utilizes a sliding window on each graph partition rather than many static snapshots. While the sliding window method focuses on processing smaller, overlapping segments of data in a sequential manner, index-batching dynamically constructs batches at runtime using indices, thus eliminating the need to store redundant data across windows and allowing workers to process disjoint segments of data. Instead of graph partitioning, we analyze the ST-GNN workflow to pinpoint memory bottlenecks and propose solutions that keep graphs intact without added communication overhead. Further, we provide a more expansive scaling study, testing with up to 128 GPUs across 32 nodes compared with up to 16 GPUs across 8 nodes. Furthermore, DynaGraph lacks a publicly available implementation, precluding direct performance comparisons or widespread community adoption.

\section{Conclusion}
In this work we explored best practices for leveraging modern HPC infrastructure in the ST-GNN training workflow. We introduced index-batching, which significantly reduces the memory footprint of single-GPU training without impacting accuracy or runtime. We further optimized index-batching as GPU-index-batching -- a method that performs the workflow entirely in GPU memory and lowers memory usage and runtime by 60.30\% and 12.87\%, respectively, for single-GPU training using the PeMS dataset. We evaluated distributed-index-batching in a distributed environment with up to 128 GPUs, displaying favorable scaling and outperforming DDP by up to 11.78x. Additionally, we proposed an extension to index-batching that utilizes optimized distributed data management techniques to reduce memory usage and epoch runtime by up to 9.00x and 2.28x, respectively, over DDP. We merge our techniques into the state-of-the-art ST-GNN package\footnote{\url{https://github.com/benedekrozemberczki/pytorch_geometric_temporal}} and publish our source code to promote reproducibility.\footnote{\url{https://github.com/uw-mad-dash/PGT_Index}} 

Building on this work, we plan to extend PGT-I to support additional spatiotemporal data structures such as dynamic graphs with temporal signal~\cite{rozemberczki2021_pytorch}. Future research might also explore options to further enhance the efficiency of distributed ST-GNN training. One option would be to investigate the integration of index-batching with graph partitioning, potentially yielding further speedups at a potential cost to accuracy. Another option is to explore data distribution strategies that allow a higher degree of user control and implement prefetching. This could help reduce the communication overhead of the distributed strategies. Additionally, by applying the tools we publish, researchers can train ST-GNNs with larger datasets more efficiently, allowing for less time-intensive application development. \footnote{ChatGPT \cite{openai2024chatgpt} and Grammarly \cite{grammarly2024} were used to improve the grammar and phrasing of this work.}

\begin{acks}
  
  This material is based upon work supported by the U.S. Department of Energy (DOE), Office of Science, Office of Advanced Scientific Computing Research and the DOE SciDAC program  under grant 0000269227 and contracts DE-AC02-06CH11357, DE-AC02-05CH11231, and DESC0012704. Additionally, this work is supported by NNSA under contract DE-NA0003525.
\end{acks}

\balance

\bibliographystyle{ACM-Reference-Format}
\bibliography{main}

\end{document}